\documentclass[10pt, conference, compsocconf]{IEEEtran}
\usepackage{makeidx}  % allows for indexgeneration
\usepackage{cite}
\usepackage{booktabs}
\usepackage[pdftex]{graphicx}
\usepackage[cmex10]{amsmath}
\usepackage{url}
\usepackage{color}
\usepackage[bookmarks]{hyperref}
\makeindex          % be prepared for an author index
\begin{document}
%
%\mainmatter              % start of the contribution
%
\title{SocioAware Content Distribution using P2P solutions in Hybrid Networks}
\author{\IEEEauthorblockN{Bernd Klasen}
\IEEEauthorblockA{SES ASTRA TechCom, Betzdorf, Luxembourg\\
Email: research@berndklasen.de}}

\maketitle              % typeset the title of the contribution
% \index{Ekeland, Ivar} % entries for the author index
% \index{Temam, Roger}  % of the whole volume
% \index{Dean, Jeffrey}
%
\begin{abstract}        % give a summary of your paper
The growing online traffic that is bringing the infrastructure to its limits induces an urgent demand for an efficient content delivery model. Capitalizing social networks and using advanced delivery networks potentially can help to solve this problem. However, due to the complex nature of the involved networks such a model is difficult to assess. In this paper we use a simulative approach to analyze how the SatTorrent P2P protocol supported by social networks can improve content delivery by means of reduced download duration and traffic.
\end{abstract}
%
% For peer review papers, you can put extra information on the cover
% page as needed:
% \ifCLASSOPTIONpeerreview
% \begin{center} \bfseries EDICS Category: 3-BBND \end{center}
% \fi
%
% For peerreview papers, this IEEEtran command inserts a page break and
% creates the second title. It will be ignored for other modes.
\IEEEpeerreviewmaketitle

\section{Introduction}\label{sec:intro}
The amount of bytes downloaded by Internet users has been steadily increasing since the birth of the world wide web and an end of this trend is not in sight. The reasons for this growth are manifold. Of course the pure number of rising Internet users is a crucial factor. Further we observe a growing number of available content, which also leads to an increase in the number of downloaded files. The effect of both---increasing numbers of files and users---is amplified by the growing file sizes. This especially applies for images and multimedia files. But these changes alone can neither sufficiently explain the characteristics of today's Internet traffic nor incidents like memes or the high speed of information spread. These processes can only be explained by further considering the paradigm change that is commonly referred to as the \textsl{web 2.0} and the appearance of online social networks (OSNs), which are causative for the dramatic speedup in observable spread of content respectively of links to it.\\

A comprehensive understanding of the origins of such events and their consequences is crucial in order to optimize the network and content delivery models. Research on the spread of information in social networks became much more comprehensive with the advent of online social networks (OSNs) and platforms like Twitter, Facebook, Google+, MySpace. Not only do these OSNs enable researches to investigate the structures of social networks in great detail, they further give insight into the personal preferences and interests of users and groups. Especially the findings on group behavior and mutual influence of individuals in social networks can be capitalized in many applications. Examples are provided in the work of \cite{Pastor-Satorras2000}, \cite{Willinger}, \cite{Scellato2011-1}, \cite{Galuba2010} and \cite{Hui2009}. In this paper, we utilize this knowledge in order to enhance SatTorrent, a P2P content distribution protocol for hybrid networks consisting of a unicast and a broadcast network. SatTorrent uses both in parallel and thus can run both with an improved efficiency, as it is described in \cite{Klasen2012-1}. The concept, preconditions, purpose and application examples are presented and analyzed in previous work \cite{Klasen2011-2}\cite{klasen2012-2}. However, the beneficial effect of capitalizing social network structures for content distribution by means of the SatTorrent protocol is still to be investigated. This is the purpose of this paper. We start with a description of the Social SatTorrent (SST) model in the following section. Later in section \ref{sec:eval} a simulative evaluation is presented that analyzes the potential of SST to improve content delivery.
%
%%%%%%%%%%%%%%%%%%%%%%%%%%%%%%%%%%%%%%%%%%%%%%%%%%%%%%%%%%%%%%%%%%%%%%%%%
%%%%%%%%%%%%%%%%%%%%%%%%%%%%%%%%%%%%%%%%%%%%%%%%%%%%%%%%%%%%%%%%%%%%%%%%%
\section{Social SatTorrent}
SatTorrent is a peer-to-peer file sharing protocol based on BitTorrent. It integrates satellite broadcasts into the delivery network in order to reduce Internet traffic by means of broadcasting popular content while unicasting the less frequently requested files. SatTorrent broadcasts payload data as well as P2P metadata files. The latter include torrent files and particularly information from trackers to peers. Since for a satellite broadcast the bandwidth demand is independent from the number of recipients, these tracker response messages can be enhanced with additional information that endows peers with a global knowledge about the overlay network which in turn facilitates an enhanced selection of exchange partners. The results of \cite{Klasen2011-2} show that we observe a significant number of publicly shared files---in that case videos---which exhibit a high number of concurrent downloaders thus making a broadcast economically reasonable. However it is worth noting that in many cases---although the absolute number of file downloads is high---the requests are more uniformly distributed over a long period of time which would hamper an efficient broadcast. In such cases, in order to relax the demand for correlation of file demand and broadcast event, clients will cache broadcasted data in case the corresponding file matches the preferences of the user. \\

As already pointed out in section \ref{sec:intro} the potential of this approach has been already evaluated. However, the analysis of SatTorrent revealed potential improvements that could be achieved from harnessing social network structures. In the remainder of this section the newly developed features that capitalize social networks in order to improve SatTorrent's performance are described. For these extensions we assume that all peers are provided with information about the user preferences and downloads of their immediate neighborhood in the social graph. Between these direct neighbors---often referred to as \textsl{friends}---an increased level of trust is assumed. That allows the sharing of this potentially critical information among them which peers might not disclose to foreign nodes. We refer to the set of nodes that have a common edge with node $n$---the direct neighbors---as the \textsl{buddies} or the \textsl{buddy-list} of this node ($B(n)$). SST does explicitly not require that all involved peers are equipped with satellite reception hardware---we refer to those peers as \textit{Sat-Peers} and to the corresponding capability as being \textit{sat-enabled}---since this can hardly be expected in a realistic scenario. Fortunately, due to the social network related protocol features also the download performance of peers that are not sat-enabled is improved, as the evaluation in section \ref{sec:eval} will show. In the remainder of this section we will introduce the social network exploitation in more detail.
%
%%%%%%%%%%%%%%%%%%%%%%%%%%%%%%%%%%%%
\subsection{Social-Network Features}
In the field of social networks and their exploitation for other services---and in particular for content distribution in peer-to-peer networks---a lot of valuable research has already been performed by numerous scientists. Hence rather than inventing the wheel from scratch SatTorrent reuses and extends known approaches. Namely these are \textit{BuddyCast} and \textit{2Fast}, which are described in the next two paragraphs before the newly developed SST features are introduced.%
\subsubsection{BuddyBroadCast}
One example of a very promising way of using social connections to increase P2P performance has been presented in \cite{Pouwelse2008Tribler}. There the authors describe the exchange of preference lists among peers in Tribler, referred to as BuddyCast. While that approach already results in a significant performance increase, it can be further optimized when a hybrid network infrastructure is available. BuddyCast periodically selects a peer to exchange preference lists with. The reason for just passing this information to one peer is mainly to keep the number of additional messages within certain limits. However, this leads to a very slow propagation of the corresponding information through the overlay network. At the same cost as one of these periodical BuddyCast messages---with respect to the message complexity---peers can send this information to a satellite uplink station. There all this information can be aggregated and---after all redundancies have been removed---be broadcasted. This is the approach taken in SST. The great benefit is that this way a vast number of peers is immediately informed  instead of letting information slowly propagate through the network. Thus peers are able to identify like-minded peers with greater accuracy and thus it further improves the performance gain of social network integration since such peers have potentially more files to exchange than those who have nothing in common. %
\subsubsection{Helping friends}
Another approach, the \textit{2Fast protocol} for collaborative P2P downloading where groups of trusted peers (buddies) denote their idle bandwidth to speed up downloads of (social) group members, is presented by the authors of \cite{Garbacki2006-2fast}. They argue that due to the asymmetric Internet connections of most users and due to the tit-for-tat policy of BitTorrent, peers could not saturate their download capacity, but it was limited by the upload bandwidth. Thus in 2Fast they allow peers to ask their \textit{buddies} for support, who will then start downloading the corresponding file from other peers and send missing pieces to the inquirer. However, half of the helpers upload capacity is used in order to receive file pieces from other peers. Nevertheless it was shown that this strategy results in a significantly increased download speed. Due to the benefits of this concept, for SatTorrent a similar approach is taken which extends 2Fast and modifies it to efficiently utilize the special conditions of the underlying hybrid network. Precisely sat-enabled peers store files that are downloaded by their friends when they are broadcasted and can then send missing pieces with their full upload bandwidth to the respective friend. We can easily see that in best case SST only needs half the helpers that 2Fast needs to saturate the Internet connection downlink bandwidth of the downloading peer since they do not need to spend upload capacity in order to get file pieces. %
\subsubsection{Persistent global upload rewarding}
One of the most important features in BitTorrent that helped to make it so successful in the tit-for-tat policy. However, for reasons of anonymity and simplicity its scope is limited to one torrent. Thus, no matter how many files a peer might have seeded before, for a new download its score starts at zero. For SST we introduce a persistent rewarding scheme based on credits. This is possible since the social network integration allows a recognition of users beyond the scope of one isolated file download and thus---in contrast to BitTorrent---a broader ambit for the tit-for-tat policy. Obviously, the requirement for user registration makes Social SatTorrent unqualified for illegal downloads. However, we do not consider this as a restriction since such behavior is undesirable anyway. The rewarding is based on credits. When peer \textit{A} requests a piece of a file from a peer \textit{B} that is not on his Buddy-List, it will try to send a piece in exchange. If \textit{B} does not need any of \textit{A}'s pieces, \textit{A} can give a credit to \textit{B} in order to satisfy the fairness policy. In order to avoid ostracism of new users, peers are allowed to have negative credits up to a certain amount specified by the global parameter \textit{CreditLimit} whose optimal value depends on the actual network size. This concept facilitates another element of SatTorrent, the donation of credits to other peers---predominantly to buddies---in order to provide download support as an addition to the approach taken by 2Fast described above. Further it gives an incentive for seeding as a source of credits.%
%
%
%%%%%%%%%%%%%%%%%%%%%%%%%%%%%%%%%%%%
\subsubsection{Social Prefetching}\label{sec:socialprefetching}
Social SatTorrent implements a prefetching strategy that triggers the download of potentially interesting files when a peer is idle. Which files might be of interest in the future is decided by examining the peer's profile, respectively its preferences. For an improved prediction of future demand the preferences and available ratings of buddies are analyzed. According to \cite{Hui2009} similar peers have a significant influence on each other. In addition there is an increased probability for buddies to exhibit similarity to a high degree. This leads to a twofold benefit: First the resulting improved reliability of predictions increases the probability for the caching peer to find a future download in its local cache already. Secondly this particular file will potentially be downloaded also by friends, which will allow the caching peer to support their download attempt. This terrestrial prefetching is a feature that all peers can use, not only those who are sat-enabled. However, the Sat-Peers additionally receive files via broadcast. On the one hand this enables them to have their caches filled much faster and with more up-to-date files. On the other hand it can reduce their unicast prefetching traffic. Besides caching files for their personal demand, Sat-Peers can donate resources for caching files that do not match their own preferences but those of their buddies. This makes sat-enabled peers a valuable resource within the network. When the caches are full, peers first replace files that are the least interesting for them. Thus files cached exclusively for friends will be removed for personal interest. However, due to the previously described affinity of closely linked individuals, personal and buddies interest should be similar to a high degree.\\
The peers that are used as sources for prefetching are limited to those who are in a nodes buddy-list. According to the results of \cite{Scellato2011-1}, \cite{Backstrom2010} and \cite{Scellato2011}---who show that the majority of social connections exhibit a small geographical distance---this causes the prefetching to happen mainly in the near physical proximity of the peer. This comes with the great advantage that the major part of the induced traffic is kept within a comparatively small geographical region and thus potentially within a single Internet service provider's (ISP) network. This is supposed to increase the acceptance of SatTorrent among the ISPs, who commonly dislike P2P file exchange protocols due to the costly inter-ISP traffic their applications produce. Further, the probability for finding a file which is a potential subject for prefetching in the buddies' caches is increased, due to the mutual influence and the increased probability for similar interests between friends.
%%%%%%%%%%%%%%%%%%%%%%%%%%%%%%%%%%%%%%%%%%%%%%%%%%%%%%%%%%%%%%%%%%%%%%%%%
%%%%%%%%%%%%%%%%%%%%%%%%%%%%%%%%%%%%%%%%%%%%%%%%%%%%%%%%%%%%%%%%%%%%%%%%%
\section{Simulating SST}\label{sec:simulation}
Since a real world test with a sufficiently large number of clients is not possible for our experiments, a simulation is utilized. In order to analyze the social network protocol features of SatTorrent, a social graph must be created. Research on social graph modeling algorithms identified several different strategies, some which are meant to be general applicable, others only for highly specialized scenarios. Among all these, the Barabasi-Albert model has been widely investigated and applied in several studies. This broad experience and validation are causal for the decision to utilize this model for the present study. However, for a better reliability of the results and in order to achieve independence from one specific model, also the Toivonen model---which has been introduced in \cite{Toivonen2006}---is implemented in the simulation. A summarization of both models is given below, together with a reasoning why they have been selected for this simulation.
%
%%%%%%%%%%%%%%%%%%%%%%%%%%%%%%%%%%%%
\subsection{Social Graph Models}\label{sec:graphmodels}
\subsubsection{Barabasi-Albert Model}\label{sec:barabasialbert}
The key strategy of the Barabasi-Albert model (BA) \cite{Barabasi2000} is preferential attachment. Starting from an initial graph consisting of a very small number of nodes ($m_0\geq2$), new nodes are added subsequently with a specified number of edges ($m$). The probability for connection to an existing node $N$ is proportional to the degree of that target node. This leads to the typical power law degree distribution which exhibits few super hubs with an extraordinary high number of connections.\\
%
%%%%%%%%%%%%%%%%%%%%%%%%%%%%%%%%%%%%
\subsubsection{Toivonen Model}\label{sec:toivonen}
In the Toivonen model (TO) new nodes are added to the graph by first connecting to $r$ randomly chosen vertices and then connecting to averagely $p$ neighbors of each of them. This results in a realistic social network structure that exhibits small world properties such as a power law distribution and short average paths as well as a high average clustering. However, it prohibits the emergence of nodes with extremely high degrees that we would find in other models, e.g. in BA graphs. These highly connected hubs are usually institutions rather than individuals. Since excluding institutional sources might be interesting for certain P2P scenarios, this specific graph model has been chosen as the alternative model.\\
%
%%%%%%%%%%%%%%%%%%%%%%%%%%%%%%%%%%%%
\subsection{Mutual influence model}\label{sec:sim:mutualinflunecemodels}
Among social scientists there is no doubt about the mutual influence between individuals in general and in particular between users in online social networks. While several studies analyze these incidents and develop different models to describe them, there is no model that can be considered as the ultimate characterization of the real world procedures. Considering this in combination with the finiteness of resources for a simulation based analysis, a decision in favor of a simple approach---in terms of the computational complexity---has been taken. In order to achieve a broader view and to allow a comparative analysis, various models have been implemented which are described in the following. For all of them the mutual influence model for every node is applied with a probability $p_MI$ which is controlled via settings file. In order to facilitate mutual influence, each user $n_i$ has a list of preferences $P(n_i)$. This list contains entries which consist of an integer that unambiguously identifies the category $C$ and a quantifier $Q(C, n_i)$ that specifies the importance of this category for the corresponding node $n_i$. Thereby, a higher quantifier indicates a higher interest in a category, with $0 < Q(C, n_i) \leq 1$.\\ 
%
%%%%%%%%%%%%%%%%%%%%%%%%%%%%%%%%%%%%
\subsubsection{Three Most significant (MI1)} \label{sec:socialnetworking:mutualmostsignificant}
The preferences of all direct neighbors $b_i \in B(n_k)$ are analyzed for each node $n_k$. Thereby the frequency of occurrences $F(C,n_k)$ of each interest category $C$ is counted and the corresponding quantifiers are summed up in
\begin
	{equation}Q_{sum}(C,n_k) = \sum\limits_{i=0}^{|B(n_k)|-1} {Q(C,b_i)}
\end{equation}
 Then one among the three categories with the highest values for the aggregated quantifiers is selected, with the probability being proportional to the number of occurrences. In case the selected user does not have this category in his preferences, it is added with the mean quantifier. Otherwise, the new quantifier $Q'(C,n_k)$  of peer $n_k$ for category $C$ is determined by the following formula:
\begin{equation}\label{eq:calculatenewquantifier}
Q'(C,n_k) = Q(C,n_k) + \frac{Q_{sum}(C,n_k)}{F(C,n_k)} \cdot  (1-Q(C,n_k))
\end{equation}
%
%%%%%%%%%%%%%%%%%%%%%%%%%%%%%%%%%%%%
\subsubsection{Three Most Frequent (MI2)}
This approach is similar to MI1 except that the criterion to select the three candidate categories is not the quantifier but the absolute occurrences of the specific categories in the neighbors' preferences. The selection of the influencing category among these candidates as well as the application of the mutual influence follows the same rules as for MI1. \\
%
%%%%%%%%%%%%%%%%%%%%%%%%%%%%%%%%%%%%
\subsubsection{Most significant only (MI3)}
For every node $n_k$ the list of neighbors $B(n_k)$ is looked up for the preference category with the highest quantifier $Q_{max}$, which is then used to influence $n_k$. In case there are  several categories which exhibit the same highest quantifier, one among them is randomly selected. In case $n_k$ does not have that category in its preferences yet, it is added with $Q(C,n_k) = 0.5 \cdot Q_{max}$. Otherwise, the modification of corresponding quantifier follows equation \ref{eq:calculatenewquantifier} with $Q_{sum} = Q_{max}$.\\
%
%%%%%%%%%%%%%%%%%%%%%%%%%%%%%%%%%%%%
\subsubsection{Random (MI4)}
From $B(n_k)$ one neighbor $b_i$ is selected randomly. Then again a random selection is applied to pick one it this nodes preference categories $C$, which is then used for influencing $n_k$. In case $C \in P(n_k)$ the new quantifier $Q'(C,n_k)$ is determined by
\begin{equation}\label{eq:calculatenewquantifierrandom}
Q'(C,n_k) = Q(C,n_k) + \frac{Q(C,b_i)}{F(C,n_k)} \cdot  (1-Q(C,n_k))
\end{equation}
and by $Q'(C,n_k) = 0.5 \cdot Q(C,b_i)$ otherwise. \\

Despite the mutual influence between individuals the opinion of users is also influenced by personal experiences. Thus whenever a user completes the download of a file in our simulation it will influence the quantifier of the corresponding category in his preferences. With a defined probability (simulation settings) this change will be negative. The magnitude of the modification is depending on the variety of interests and the current quantifier value. 
%
%%%%%%%%%%%%%%%%%%%%%%%%%%%%%%%%%%%%
\subsection{Static Simulation Parameters}
For the sake of comparability as many parameters as possible are kept unchanged throughout the evaluation phase. These are summarized in table \ref{tab:simparameters}. Each of the 10 initial seeders provides all files. All participants---including the seeders---share the same asymmetric connection bandwidth. After a peer finished a download, before starting a new one it waits for a random time whose global average is set to two hours. The remaining parameters are introduced later in section \ref{sec:eval} with their particular values. %
\begin{table}[t]
\centering
 \sffamily
\begin{tabular}{p{3.7cm}r}
\toprule 
\textbf{Parameter} & \textbf{value} \\
\midrule 
File Size & 100 MB \\ 
Categories & 100 \\ 
Seeders (initially) & 10 \\ 
Download Bandw. & 8 Mbit \\
Upload Bandw. & 1 Mbit \\
Sat-Enabled Users & 30\% \\
Avg. download wait time & 2h \\
\bottomrule 
\end{tabular} 
\caption{Static Simulation Param.}
\label{tab:simparameters}
\end{table}

%
%%%%%%%%%%%%%%%%%%%%%%%%%%%%%%%%%%%%%%%%%%%%%%%%%%%%%%%%%%%%%%%%%%%%%%%%%
%%%%%%%%%%%%%%%%%%%%%%%%%%%%%%%%%%%%%%%%%%%%%%%%%%%%%%%%%%%%%%%%%%%%%%%%%
\section{Evaluation}\label{sec:eval}
%
%%%%%%%%%%%%%%%%%%%%%%%%%%%%%%%%%%%%
\subsection{Dish Distribution Analysis}\label{sec:eval:dishdensity}
\begin{figure}
\centering
		\includegraphics[width=0.8\columnwidth]{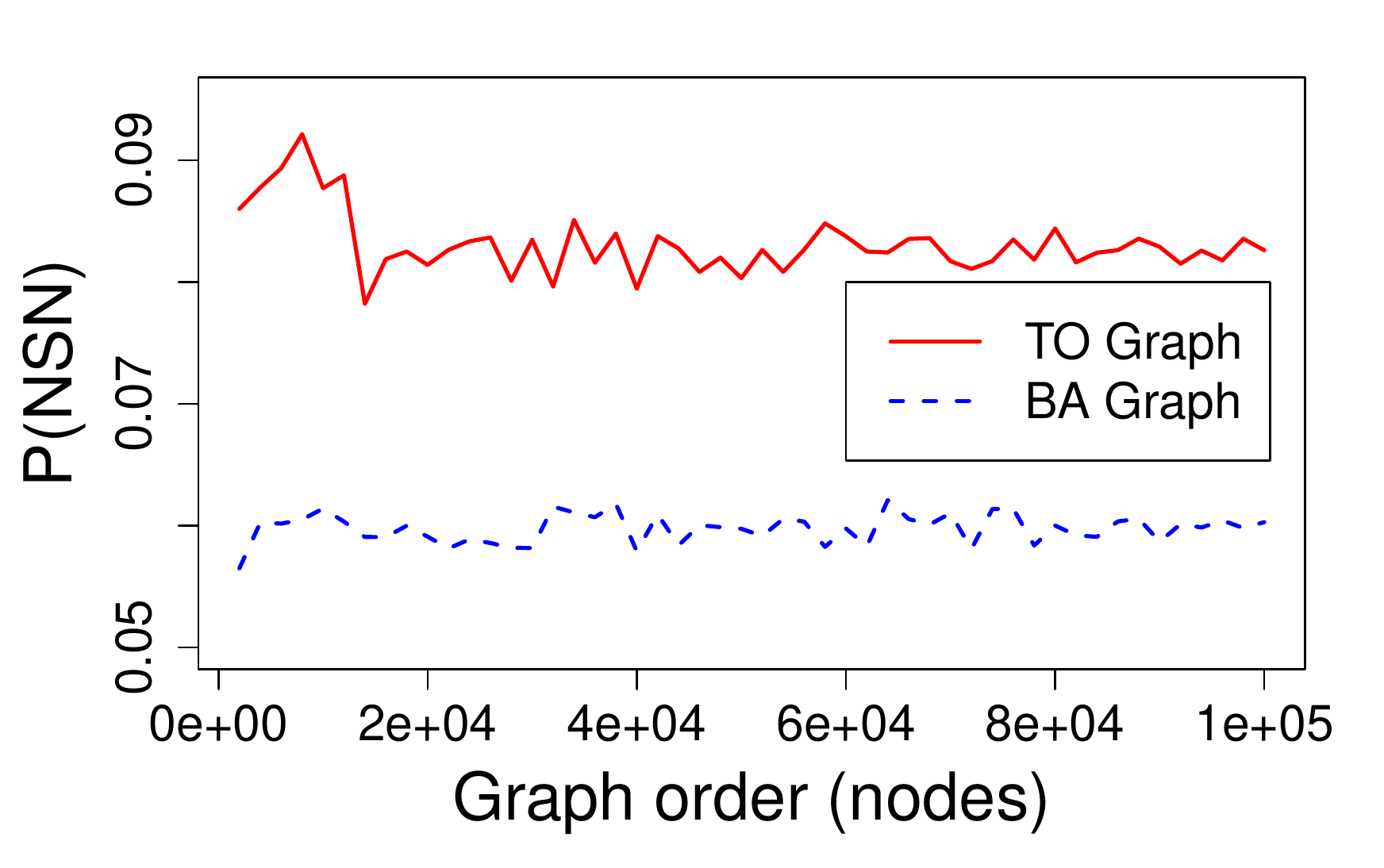}
		\caption{$P(NSN)$ for varying graph sizes}
		\label{fig:ddnodes}
\end{figure}

As results of preceding SatTorrent evaluations show, the protocol's performance increases with higher ratios of sat-enabled peers at least up to a fraction of 95\% of sat-enabled peers \cite{Klasen2012-1}. In this %
\begin{figure}[b]
	\centering
		\includegraphics[width=0.8\columnwidth]{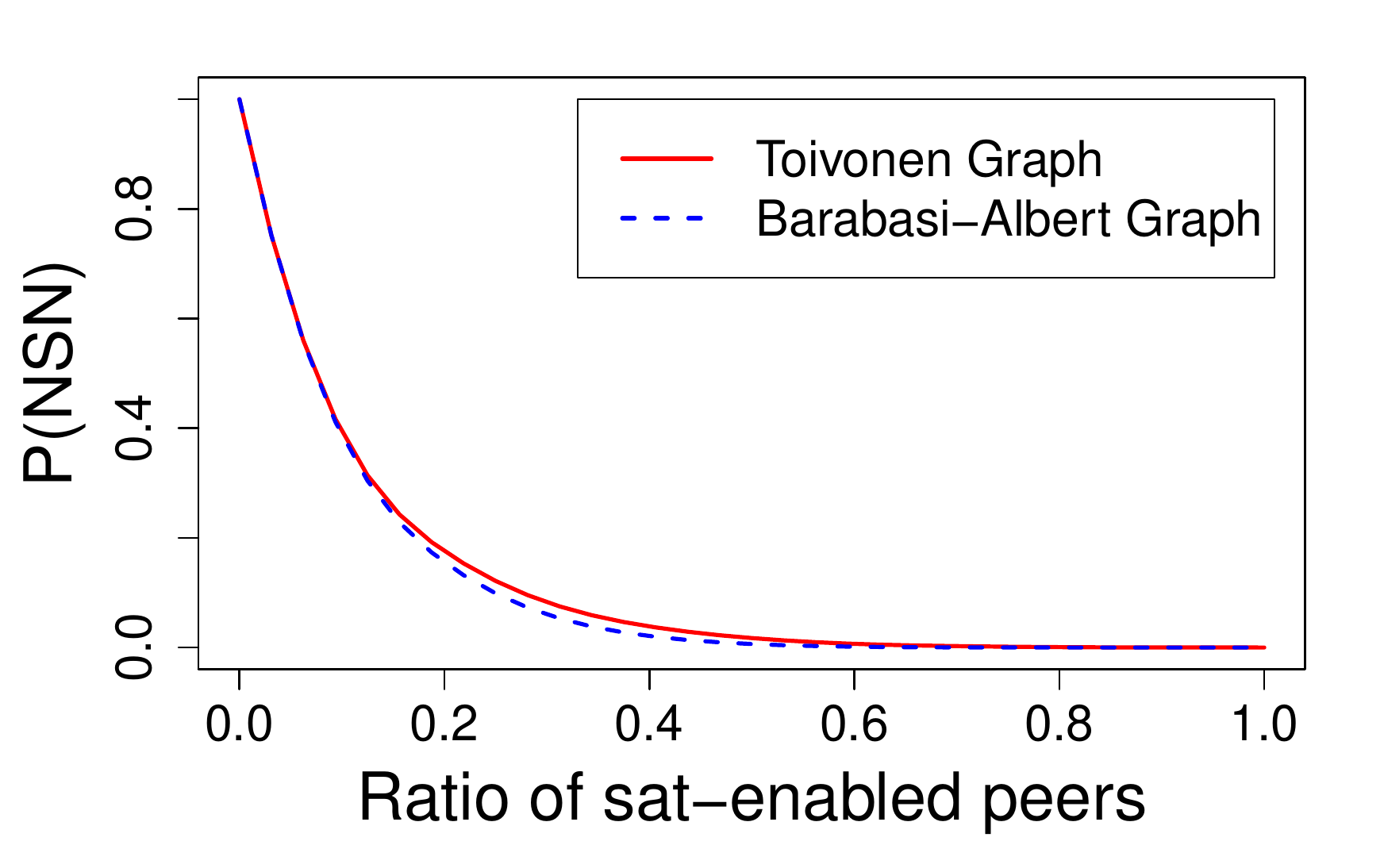}
		\caption{$P(NSN)$ for varying dish ratios}
		\label{fig:dddishdensity}
\end{figure} %
context we analyze the distribution of sat-enabled peers across social network communities. Therefore the social network graphs are generated according to the models introduced in section \ref{sec:graphmodels} and analyzed afterwards. As we already discussed above, SST not necessarily demands for each peer having a satellite reception unit. More important is the aspect of having a neighbor which is sat-enabled. Thus we analyze the probability of peers for having no sat-enabled neighbor ($P(NSN)$). First this probability is examined under different proportions of sat-enabled peers, varying from 0\% up to 100\%. The results are visualized in figure \ref{fig:dddishdensity}. Both curves are rapidly decreasing, while in the BA graph the probability recedes slightly faster. This small but steady difference can also be observed in figure \ref{fig:ddnodes} where we compare $P(NSN)$ under varying numbers of nodes and a constant ratio of sat-enabled peers of $0.3$. No significant influence of the graph size on $P(NSN)$ can be recognized. Further, the difference between the BA and the TO is obvious but very small.\\
However, the differences in the graph structure are distinctive. In table \ref{tab:graphproperties} we find a comparison of the most important properties of each graph type. For this analysis a graph size of 10,000 nodes and a Sat-Peer ration of $0.3$ were used for both graph types. Especially the average clustering coefficients differ significantly. Whether or not this has a considerable impact on the performance of SST is analyzed in section \ref{sec:eval:cdefficiency} after an investigation on the impact of different mutual influence models in following section.
\begin{table}[b]
\centering
 \sffamily
\begin{tabular}{p{3.8cm}rr}
\toprule 
 & \textbf{BA} & \textbf{TO} \\
\midrule 
Nodes & 10,000 & 10,000 \\ 
Edges & 49,985 & 50,921 \\ 
Average degree & 9.997 & 10.184 \\ 
Network diameter & 6 & 9 \\ 
Avg. clust. coeff. & 0.007 & 0.527 \\ 
Avg. path length & 3.655 & 4.328 \\ 
Total triangles & 2,194 & 61,904 \\ 
\bottomrule 
\end{tabular} 
\caption{Comparison of graph properties}
\label{tab:graphproperties}
\end{table}
%
%%%%%%%%%%%%%%%%%%%%%%%%%%%%%%%%%%%%
\subsection{Impact of Mutual Influence Model}\label{sec:eval:mutualinfluence}
\begin{figure}[t]
	\centering
		\includegraphics[width=0.8\columnwidth]{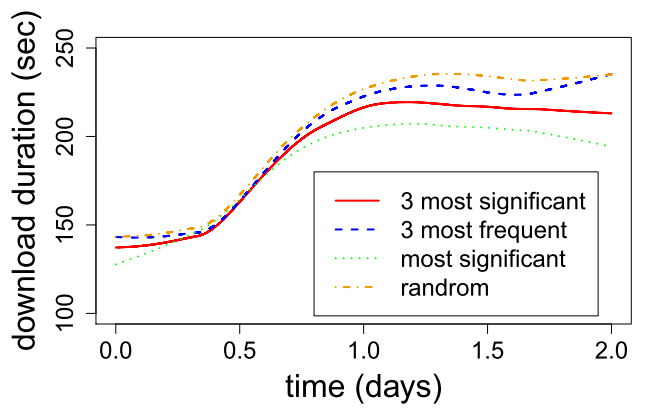}
		\caption{Download duration for varying MIs}
		\label{fig:mi:duration}
\end{figure}

In section \ref{sec:sim:mutualinflunecemodels} we introduced different ways of modeling the mutual influence among users. Since the prediction of future user behavior and thus the prefetching, caching and cache replacement strategies rely on the analysis of the user profiles and those of the corresponding buddies, the mutual influence model is crucial for the reliability of the simulation results. However, it is impossible to define one universal mutual influence model that would correctly reflect the behavior for all possible applications of SST. The comparison provided in this section will show how much the performance of SST is affected by variation in this model. In this context we first compare the average download duration for the different models. Besides three realistic models (MI1, MI2, MI3) we do this analysis also for a fourth, random approach (MI4). The latter is expected to deliver rather undesired results since it should lead to cache misses and failed prefetching attempts. Figure \ref{fig:mi:duration} %
shows a comparison of average download times. While all the four models are close together, we indeed observe the worst results for the random approach. The common increase in the download times after 12 hours is caused by the increasing number of concurrent downloads paired with a high prefetching activity. Thus we reach a point where insufficient sources are available to satisfy the demand. We recall, with all peers having a asymmetric Internet connection with an download/upload ration of 8:1, it is simply impossible to have all peers concurrently downloading with full speed. Whether or not this means that SST with prefetching leads to a better or maybe worse overall performance than non-social SatTorrent or other P2P protocols will be investigated later in section \ref{sec:eval:cdefficiency}.\\

Another important factor is the number of bytes that are exchanged between peers that are not neighbors in the social graph, since we assume this traffic to be more costly for ISPs. The corresponding results are visualized in %
\begin{figure}[t]
	\centering
		\includegraphics[width=0.8\columnwidth]{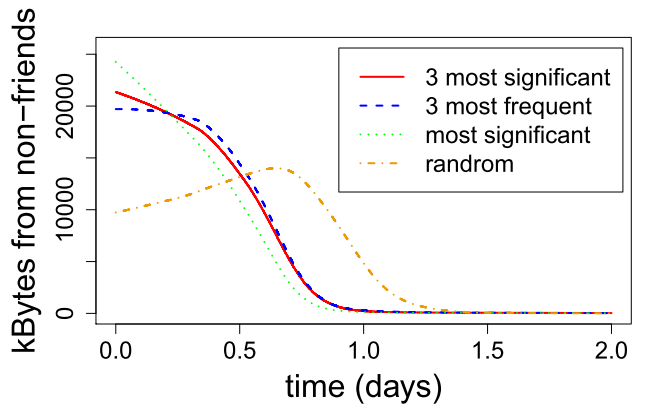}
		\caption{Non-friend upload comparison}
		\label{fig:mi:foreignpieces}
\end{figure}%
figure \ref{fig:mi:foreignpieces}. While MI1-3 again show only very little difference, the random approach shows a different behavior with lower values at the start. The reason is that at the beginning the difference between the neighbors are still more significant, while they assimilate more and more in the course of the simulation. Thus a random approach can perform better. Interestingly, all curves converge to zero after a relatively short period. This is a result of the satellite broadcasts and the prefetching. The effects of these features are investigated in more detail in section \ref{sec:eval:cdefficiency}.\\
\begin{table}[b]
	\centering
\begin{tabular}{p{2cm}ll}
	\toprule 
 & \textbf{Sat-Peer} & \textbf{\#Sat-Peer Friends} \\
	\midrule 
3 most significant (MI1) & 0.4613411 & -0.1048858 \\ 
3 most frequent (MI2) & 0.499472 & -0.1194338 \\ 
most significant (MI3) & 0.3898186 &-0.09718519 \\ 
random & 0.5298729 (MI4) & -0.1245606 \\ 
	\bottomrule 
\end{tabular} 
\caption{Comparison of correlation coefficients}
	\label{tab:mi:corcoeff}
\end{table} %
Regarding the different mutual influence models we observe slightly better results for MI3. When we have a look at the correlation coefficients of download duration and being sat-enabled respectively the number of sat-enabled friends (table \ref{tab:mi:corcoeff}), we get a better understanding of this difference. For MI3, having a dish is important, but not as much as for the other models. The same relation is observed for the number of sat-enabled friends. The reason is that MI3 leads to user preferences where some categories have a very high rating while others remain low. Thus less files are sufficient for a good cache hit rate. What can be derived from this is that the less exactly the future demand can be predicted, the more important the satellite broadcasts are. 
\subsection{Content Distribution Efficiency}\label{sec:eval:cdefficiency}
\begin{figure}[!t]
	\centering
	\begin{minipage}[b]{0.45\columnwidth}
		\includegraphics[width=\columnwidth]{d2-dl-duration.png}
		\caption{Comp. of download \newline durations}
		\label{fig:cd:duration}
		\end{minipage}
	\begin{minipage}[b]{0.45\columnwidth}
		\centering
			\includegraphics[width=\columnwidth]{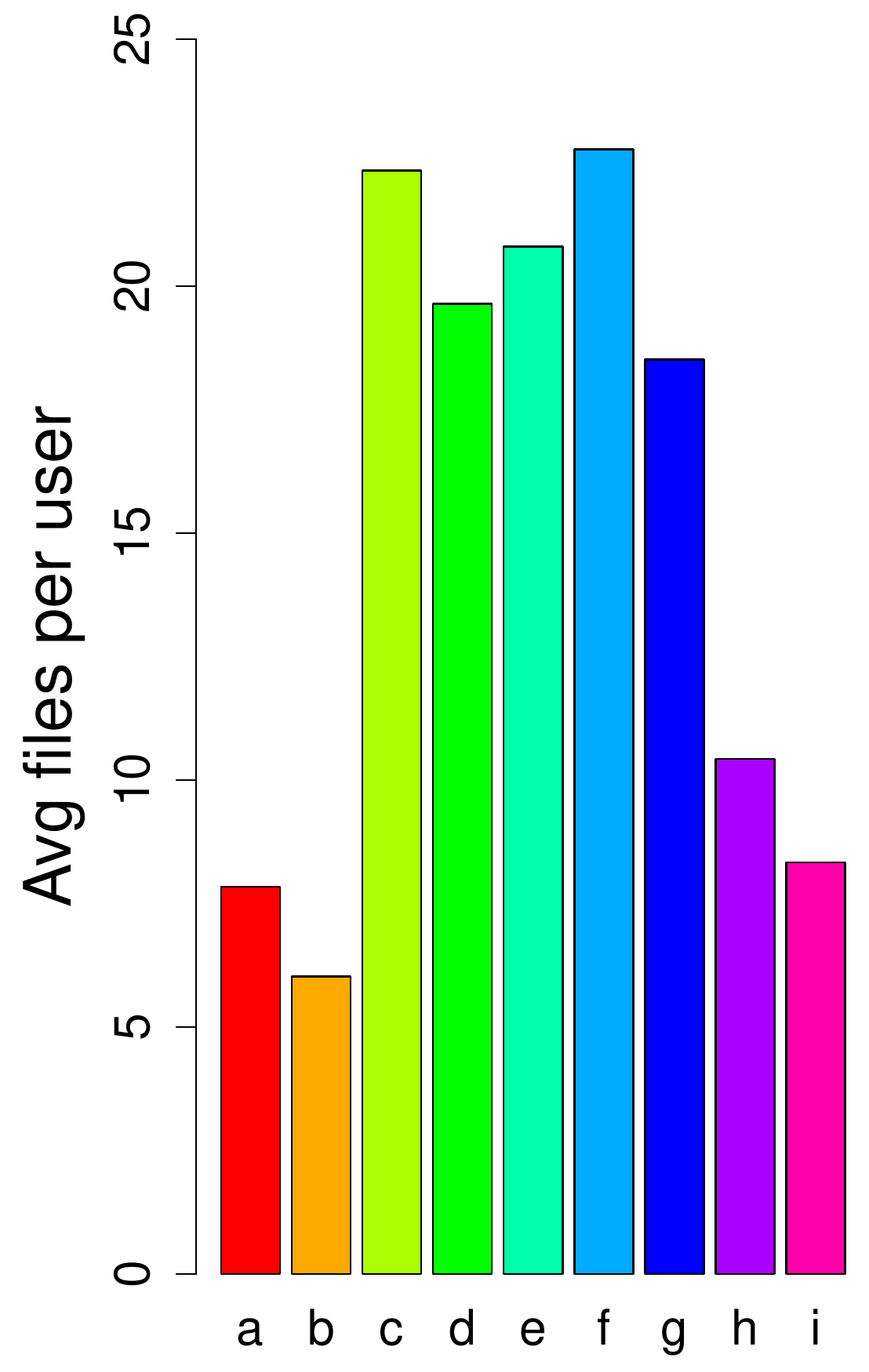}
		\caption{Files downloaded per user at different settings}
		\label{fig:cd:totalpieces}
		\end{minipage}
\end{figure}

For the experiments in this section we utilize one satellite transponder with a bandwidth of 36 Mbit (error correction already deducted). Figure \ref{fig:cd:duration} shows the mean download duration for different settings. For all figures, the mapping of the characters is the same. For 'a' all features are switched off, which makes it comparable to usual BitTorrent. For 'b' download support from friends is enabled, 'c' comes with prefetching and for 'd' prefetching is limited to 10 concurrent users per step. 'e' to 'i' have broadcasts enabled. Except that, 'e' uses the same settings as 'd', 'f' the same as 'c', 'g' the same as 'b' and 'i' the same as 'a'. In 'h', information from the social network is used for keeping traffic local but download support from friends is deactivated. For 'a' and 'b' we observe in figure \ref{fig:cd:duration} a significantly increasing duration. This is due to a shortage of sources. The remainder all perform much better, while the broadcasting counterpart always delivers better results. The difference might look small due to the axis scale, but broadcasts reduce the mean duration by one third. The results for 'h' and 'i' are very impressive on the first view. However, figure \ref{fig:cd:totalpieces} reveals that the average number of files downloaded by each user is much lower for these settings. What happened here is that sat-enabled peers---due to cached content---often have download times of zero seconds, since the files are already completely cached when they start the corresponding download. However, non sat-enabled peers wait very long for their downloads. Since they thus do only download a small number of files, they do not affect the mean values notably which leads to the curve progression in figure \ref{fig:cd:duration}.\\
In figures \ref{fig:cd:foreignpieces1} and \ref{fig:cd:foreignpieces2} we compare the amount of data received from peers that are not in the downloader's buddy list (not friends) which is potentially expensive traffic. Figure \ref{fig:cd:foreignpieces2} shows absolute values and exhibits extraordinary high values for setting 'a'. This is the reason why its curve does not appear in figure \ref{fig:cd:foreignpieces1} where the range has been limited to the most important section. What can be observed here is that those simulations that used broadcasts converge to lower values than their non-broadcast counterparts. Considering that the absolute unicast traffic is only slightly lower for the broadcast approaches, this is an important factor. 

\section{Conclusion}
In this paper we compared SST's performance to the standard SatTorrent protocol and to non-satellite supported models. Thereby we aimed at an optimization of download duration. In all comparisons SST outperformed the other models. However, the Internet bandwidth consumption is only slightly lower. Altogether we obtain a good overall result for the SST performance, a significantly reduced average download time combined with a higher throughput, lower total Internet traffic, and a potential decrease of total long range traffic. However, there is still much potential for improvement in future protocol versions. Further SST can be adjusted differently in order to deliver better performance for usage scenarios where traffic reduction is the most important factor. The investigation of the corresponding settings is subject to future work. %

\section*{Acknowledgment}
This research is supported by the \textsl{National Research Fund} of Luxembourg.

%

% Generated by IEEEtran.bst, version: 1.13 (2008/09/30)

\end{document}